%% file: arxiv.tex
\documentclass{article}

\usepackage{arxiv}
\usepackage{lineno}
\modulolinenumbers[5]

\usepackage[utf8]{inputenc} 
\usepackage[T1]{fontenc}    
\usepackage{url}            
\usepackage{hyperref}
\usepackage{booktabs}       
\usepackage{amsfonts}       
\usepackage{nicefrac}       
\usepackage{microtype}      
\usepackage{rotating}
\usepackage{hhline}
\usepackage{booktabs,threeparttable}
\usepackage[table]{xcolor}
\usepackage{array}
\usepackage{makecell}
\usepackage{tabularx}
\usepackage{diagbox}
\usepackage{color,soul}
\usepackage{graphicx}
\usepackage{bigstrut}
\usepackage{amssymb}
\usepackage{subcaption}
\usepackage{multirow}
\usepackage{multicol}
\usepackage{natbib}

\newcommand*\rot{\rotatebox{90}}
\newlength\tbspace
\newcolumntype{L}{l<{\hspace{\tbspace}}}

\title{Mobile Social Media Usage and Academic Performance}

\author{
  Fausto Giunchiglia \\
  Department of Information Engineering \\
  and Computer Science\\
  University of Trento\\
  Trento, Italy \\
  \texttt{fausto.giunchiglia@unitn.it} \\
   \And
  Mattia Zeni \\
  Department of Information Engineering \\
  and Computer Science\\
  University of Trento\\
  Trento, Italy \\
  \texttt{mattia.zeni.1@unitn.it} \\
  \And
  Elisa Gobbi \\
  Department of Sociology and Social Research\\
  University of Trento\\
  Trento, Italy \\
  \texttt{elisa.gobbi@unitn.it} \\
  \And
  Enrico Bignotti \\
  Department of Information Engineering \\
  and Computer Science\\
  University of Trento\\
  Trento, Italy \\
  \texttt{enrico.bignotti@unitn.it} \\
  \And
  Ivano Bison \\
  Department of Sociology and Social Research\\
  University of Trento\\
  Trento, Italy \\
  \texttt{ivano.bison@unitn.it} \\
}

\begin{document}
\maketitle

\begin{abstract}
\input{abstract}
\end{abstract}


\input{introduction}
\input{soa}
\input{solution}

\input{smartunitn}

\input{results}

\input{correlation}

\input{conclusions}

\bibliographystyle{plainnat} 
\bibliography{arxiv}

\end{document}

%% file: abstract.tex
Among the general population, students are especially sensitive to social media and smartphones because of their pervasiveness. Several studies have shown that there is a negative correlation between social media and academic performance since they can lead to behaviors that hurt students' careers, e.g., addictedness. However, these studies either focus on smartphones and social media addictedness or rely on surveys, which only provide approximate estimates. We propose to bridge this gap by \textit{i)} parametrizing social media usage and academic performance, and \textit{ii)} combining smartphones and time diaries to keep track of users' activities and their smartphone interaction. We apply our solution on the 72 students participating in the SmartUnitn project, which investigates students' time management and their academic performance. By analyzing the logs of social media apps on students' smartphones while they are studying and attending lessons, and comparing them to students' credits and grades, we can provide a quantitative and qualitative estimate of negative and positive correlations. Our results show the negative impact of social media usage, distinguishing different influence patterns of social media on academic activities and underline the need to take it into account and control the smartphone usage in academic settings.

%% file: introduction.tex
\section{Introduction}

Social\stepcounter{footnote}\footnotetext{This paper is an extended version of \citep{giunchiglia2017mobile}, presented at the  9th International Conference on Social Informatics (SocInfo 2017).} media have penetrated in the everyday life of Internet users, and the increasing pervasiveness of smartphones is only strengthening this phenomenon. Nowadays, these two technologies are intertwined, since smartphones are becoming more and more pervasive, especially in the student population.
\par In the literature, there is overwhelming evidence of the negative impact of social media \citep{paul2012effect}, especially Facebook \citep{junco2012too,meier2016facebocrastination}, and smartphone usage on academic performance \citep{lepp2015relationship,samaha2016relationships}.  These technologies combined lead to a series of behaviours that cause students to dedicate more time to them than actually studying. Usually, surveys are used to identify correlations between students' behaviour and their academic performances, coupled with scales that understand different metrics of students' behaviour. However, using surveys often fails to fully capture these behaviors. In fact, \citep{leecomparing} finds that users underestimate their usage time by 40\% than reported, while \citep{boase2013measuring}, although focusing only on SMS and calls, notes that self-reports suffer from low criterion validity and lead especially to overreporting usage, in contrast with \citep{leecomparing}. Furthermore, \citep{andrews2015beyond} notes that users show great lack of awareness of the frequency with which they check their phone. On the other hand, works studying smartphone usage and social media tend to focus on addictedness \citep{leecomparing,lee2014hooked}. These studies investigate students since it is a sample of population very susceptible to smartphone penetration, but they do not attempt to correlate smartphone usage patterns to academic life. Thus there is a gap between this work on addictedness and sociological surveys on academic performance, since the former could be exploited to effectively corroborate the latter with usage logs.
\par We propose to bridge this gap via a systematic approach consisting in defining new metrics for representing social media usage and using smartphones to both track app usage and administer time diaries \citep{sorokin1939time}, a well known sociological tool for understanding people's time use. This innovative coupling allows us to isolate the time of specific activities related to academic performance and provide new insights on behavioural correlations.
\par We apply this approach on a subset of data about social media apps from the SmartUnitn project, which aims at correlating the time management of students and their academic performances --- this also covers how they manage their time when interacting with social media. We  extract social media usage from students' smartphones during specific activities related to university life, i.e., studying and attending classes, and compare it with respect to their GPA and credits obtained. Results show that there is a negative correlation between the use of social media and academic performance, with different patterns depending on the activity. 
\par The remainder of this paper is organized as follows. Section \ref{sec:soa} illustrates the state of the art, and the main issues with respect to sociological surveys and studies on addictedness. Section \ref{sec:solution} describes our proposed solution, while Section \ref{sec:smartunitn} explains the dataset it is based on, i.e., SmartUnitn. Section \ref{sec:usage} and Section \ref{sec:correlation} show our results on the correlation between social media usage and academic performance. Finally, Section \ref{sec:conclusions} concludes the paper.

%% file: soa.tex
\section{Literature Review and Hypothesis}
\label{sec:soa}
Studies using smartphones on students to understand the link between addictedness and usage usually divide them in two groups (potential addicts and non addicts) based the the Smartphone Addiction Scale \citep{kwon2013development}, which consists of ten items in a six-point Likert-type scale (1 = ``I absolutely disagree", 6 = ``I absolutely agree"). \citep{lee2014hooked} used it on 95 students and then had them install the SmartLogger software to record specific events related to users' interactions with their phones, e.g., touch, text input and active/inactive events. The extracted patterns indicate that addict risk groups tended to spend more time on their apps, focusing on those that gave them instant gratifications, e.g., entertainment. Similarly, \citep{leecomparing} had 35 students download an application that monitored their smartphone usage for 6 weeks. Results show that while messenger apps were the most used applications for both groups, addicts strongly preferred social media applications.
\par Students are also the main sample investigated in reality mining \citep{eagle2006reality}. In terms of social media, the Copenhagen Networks Study \citep{stopczynski2014measuring} is currently collecting data on 1,000 students by coupling smart phone data with face-to-face interactions and Facebook usage; however, they do not consider these data in relation to students' academic performance. In terms of academic performance, the SmartGPA study \citep{wang2015smartgpa} used data from the Student Life study \citep{wang2014studentlife}, which analyzed the impact of workload on several mental and physical aspects of students' life, e.g. mood, and sociability, of a class of 48 students across a 10 week term, to show that there is evidence of a link between the students' GPA and their behaviour. However, \citep{wang2015smartgpa} did not consider social media usage to analyze their impact on students' career, although this type of information was collected.
\par In the sociological community, studies show that there is a negative relation between social media usage and academic performance. For instance, \citep{rosen2013facebook} investigated the behaviour and settings of study for 263 students of different levels of education, i.e., middle school, high school, and university. Observers controlled students for 15 min and recorded their on-task and off-task behavior every minute. On average, students became distracted in less than 6 minutes before switching to technological distractions such as social media and texting. \citep{junco2012too} focused on how Facebook use is related to academic performance, by surveying 1839 college students on their use of Facebook and then comparing it to their GPA. The results indicate that a negative correlation exists between time spent on Facebook and GPA. Overall, it appears that social media provide students with immediate  pleasure in comparison of other activities such as studying or attending lessons \citep{jacobsen2011wired}.
\par Much like social media, using smartphones also negatively affects academic performance \citep{al2015smartphone}; indeed, social media are becoming more and more synonymous in usage with smartphones. In fact, \citep{jeong2016type} notices that social networks can be used to predict smartphone addiction in users due to their pervasiveness and connectivity. These features of smartphones lead to multitasking \citep{lepp2015relationship}, i.e., the use of social media while doing something else, which is detrimental to the time dedicated to academic activities. \citep{lepp2015relationship} conducted a survey on US college students, analyzing their notions of self efficacy and self regulation, i.e., how well they believe that they can attain their goals and how they can regulate and control themselves, when using smartphones. Those students with low self regulation turned out to be the one whose usage of smartphones affects their academic performance the most. In terms of demographics, \citep{al2015smartphone} suggests that gender and field of study may act as addiction predictors. From their review of the literature, it appears that males and humanities students tend to be more susceptible to smartphone addiction.
\par However, some research highlights how surveys used to establish these correlations may be unreliable, leading to an approximation of actual usage \citep{leecomparing,boase2013measuring,andrews2015beyond}. One reason is that surveys are based on aggregate data from `stylized" questions \citep{juster1985time}, e.g., ``How many times a day on average do you check your smartphone?" \citep{gokccearslan2016modelling}, which force users to recalling activities and finding an appropriate form of  averaging \citep{kan2008measurement}. On the other hand, works relying on smartphone data for analyzing usage tend to focus on addictedness on its own \citep{leecomparing,lee2014hooked} or do not correlate usage patterns to academic performance \citep{wang2015smartgpa}. 

%% file: solution.tex
\section{Social media usage and academic performance}
\label{sec:solution}
Our proposed solution consists of two elements: \textit{i)} defining metrics for capturing the smartphone usage patterns in terms of social media, illustrated in Section \ref{sec:parameters}, and \textit{ii)} employing together time diaries and smartphones to establish the correlation between social media usage and academic performance, described in Section \ref{sec:tools}. This second element provides the systematic aspect of our approach since it couples the two tools to overcome their respective limitations and produce more accurate results.
\subsection{Parameters for social media usage and academic performance}
\label{sec:parameters}
To represent social media usage and academic performance, we define three different parameters: \textit{i)} \textit{social media}, \textit{ii)} \textit{usage} and \textit{iii)} \textit{academic performance}.
\par \textit{Social media} (applications) are any technology used to share textual, image and audio content. We further divide social media applications, hence SM, in three categories: 

\begin{itemize}
\item \textbf{Social network sites (SNS):} online platforms used by people to build social networks or social relations with other people, e.g., Facebook;
\item \textbf{Instant messaging applications (IM):} online chats that offer real-time text transmission over the Internet, e.g., Whatsapp;
\item \textbf{Browsers (Web):} software applications for retrieving, presenting and traversing information on the Internet, e.g., Chrome
\end{itemize}

This distinction allows to capture the fact that each type requires different usage patterns and threatens students' performances accordingly \cite{junco2012too,lepp2015relationship}. For instance, people use SNS for a longer period of time than IM \cite{meier2016facebocrastination} and both negatively affect students' performance, while browsers may be used to access both  academic and non academic topics, e.g., going on Youtube for entertainment purposes vs going on Wikipedia for studying.  
\par To represent and evaluate \textit{the usage of social media}, we distinguish between three types of interactions between students and their smartphone applications:

\begin{enumerate}
\item \textbf{$\bar S$}: the average number of occurrences of social media app usage, i.e., \textit{sessions}  of students checking social media.;
\item \textbf{$\bar D$}: the average time of social media app usage (in seconds), i.e., the \textit{duration} of the social media sessions, namely where any social media app is running ;  
\item \textbf{$\bar I$}: the average time in between app usage (in seconds), namely when there is known human interaction (swiping/typing) with an app, i.e., the duration of the \textit{inactivity} of the phone 
\end{enumerate}

Notice that $\bar S$ and $\bar D$ extend and provide further structure to the notion of frequency from \cite{andrews2015beyond}, which only accounts for the former without considering its duration as parameter.
\par We represent \textit{academic performance} with two measures: 

\begin{itemize}
\item \textbf{Grade Point Average (GPA):} the average of grade points a students obtained during a semester. It represents the qualitative dimension of academic performance, since it refers to the how well students perform;
\item \textbf{Credito Formativo Universitario (CFU):} course credits obtained by students for each exam taken. They represent the quantitative dimension of academic performance, since they refer to the progress of their university career.
\end{itemize}

Additionally, socio-demographic variables must be accounted for; in this work, following \cite{al2015smartphone}, students' faculties (scientific and humanities) are treated as socio-demographic variables to predict the effect of social media on academic performance.
\subsection{Time diaries via smartphones}
\label{sec:tools}
In terms of sociological survey tools, we employ \textit{time diaries}. Time diaries are logs where respondents are asked to detail how they allocated their time during the day \cite{sorokin1939time}. These logs are tables divided in time intervals of 10 or 15 minutes \cite{romano2008time}, and provide information of time use in terms of activities performed, locations visited and people encountered during the 24-hour period \cite{hellgren2014extracting}. 
\par Although time diaries also suffer from issues of reliability as all self-reports do \cite{kan2008measurement}, they provide the following advantages:

\begin{enumerate}
   \item Since respondents have to keep a log of their activities, time diaries allow us to acquire information not only on the average amount of time spent on different activities during a day, but also the duration and frequency of each activity, together with their sequences;
   \item Time diaries provide a systematic tool for also understanding spatial and social relations of users, which enriches and widens our scope of research;
   \item Respondents do not need to provide average estimates in time diaries, which lessens the cognitive load while completing them \cite{kan2008measurement}, while also reducing the possible mismatch between these answers given and the actual usage of smartphones.
\end{enumerate}

In this work, we employ a time diary shown in previous work \cite{giunchiglia2017context}, which asks users three questions: \textit{i)} ``What are you doing?", i.e., activities like ``shopping", \textit{ii)} ``Where are you?", i.e., places like ``home", and \textit{iii)} ``Who is with you?", i.e., social relations like ``family". The possible answers are a list of pre-defined labels, which minimizes coding, adapted from the ATUS time use survey \cite{shelley2005developing}.
\par Smartphones can enhance time diaries by administering them to users, which then are able to answer them in (almost) real time, in addition to performing sensor collection, e.g., GPS, Bluetooth, call logs, and running applications, among others. These two functionalities of smartphones can be exploited to match any given triple of reported activity, location, and social relation with the status of the smartphone as a proxy of the actual user behavior. Our approach allows us to compare social media usage and academic performance,  by matching those activities that are directly linked to academic outcomes, i.e., studying and attending lessons, together with the running applications, which include SM, at the time of the answer. 



%% file: smartunitn.tex
\section{Methods}
\label{sec:smartunitn}

We validate our proposed solution on the data from the SmartUnitn project, which belongs to a family of projects called \reflectbox{S}\reflectbox{M}\reflectbox{A}\reflectbox{R}TRAMS\footnote{See http://trams.disi.unitn.it for more information} that leverages on smartphones to extract behavioural patterns from people and develop systems that assist users in their everyday life. The SmartUnitn  project aims at investigating how students' time allocations affects their academic performance. 
\par The project relies on the i-Log mobile application \citep{zeni2014multi,giunchiglia2017context} to provide the two functionalities needed from smartphones in our approach:

\begin{itemize}
    \item \textbf{Data collection:} i-Log is designed to collect data from multiple sensors simultaneously, both hardware (e.g., GPS, accelerometer, gyroscope, among others) and software (e.g., in/out calls, applications running on the device). A dedicated backend infrastructure manages the tasks of synchronizing and storing the streams of data from the smartphones. 
    \item \textbf{Time diaries:} i-Log can administer the time diary from \citep{giunchiglia2017context} as a question composed of three sub-question on activities, locations and social relations of students every 30 minutes. Every triple of questions can be answered within 150 minutes from its notification, with a maximum of 5 questions stacked in queue, otherwise it expires and treated as null. Questions appear as a silent notifications in order to avoid bothering students and disrupt their activities too much.
\end{itemize}

Based on two years of internal testing, i-Log has been designed to \textit{ii)} be modular and adapt to each smartphone model, especially in terms of sensing strategies for both  smartphones and their internal sensors (which can greatly vary among different models) \textit{ii)} to consume as little battery as possible, by devising sensor-dedicated energy consumption strategies and delegating all computation server-side, and \textit{iii)} to ensure users' privacy from data collection to its analysis. 

\subsection{Participants}
\par In this project, 72 students were selected from the ones enrolled at the University of Trento in the academic year 2015-2016 and in particular only those who fulfilled three specific criteria: \textit{i)} to have filled three university surveys in order to obtain their socio-demographic data, shown in Table \ref{tab:socio-demo}, and other characteristics, e.g., psychological and time use related; \textit{ii)} to attend lessons during the period of our project in order to describe their daily behavior during the university experience, and \textit{iii)} to have an Android smartphone with an Android version 5.0.2 or higher. 

\begin{table}[thb!]
\centering
\caption{Socio-demographics of students from our project}
\label{tab:socio-demo}
\begin{tabular}{|l|l|l|l|l|l|}
\hline
\multicolumn{2}{|c|}{\textbf{Gender}} & \multicolumn{2}{c|}{\textbf{Departments}} & \multicolumn{2}{c|}{\textbf{Scholarship}} \\ \hline
Male              & Female            & Scientific          & Humanities          & True                & False               \\ \hline
61.1\%            & 39.9\%            & 56.9\%              & 43.1\%              & 37.5\%              & 62.5\%              \\ \hline
\end{tabular}
\end{table}

\par The students were asked to attend an introductory presentation where they are presented with the aims of the project and how to use the application. If they wished to participate, after the presentation they signed a consent form, and then installed i-Log on their own smartphones. Users were informed about all aspects of the management of their personal information concerning privacy, from data collection to storage to processing. Furthermore, before starting the data collection, we obtained the approval from the ethical committee of our university.

\subsection{Data collection}
\par The project lasted two weeks. In the first one, students used i-Log to answer to time diaries while also having their data collected; during the second week, they were only required to have the application running for the collection of data.


\par The SmartUnitn dataset amounts to 110 Gb, containing behavioural data from smartphones merged with socio-demographic characteristics of students obtained both through surveys and academic performance data provided by the University of Trento. 

\subsection{Measures}
\par In terms of our parameters, the SmartUnitn dataset provides the following data:

\begin{itemize}
    \item \textbf{Social media:} There 957 different applications  across all students, and, within 32 SM apps, social networks are the most represented with 11 apps.
    \item \textbf{Usage:} Because of the Android operating system design, any application in the foreground keeps being logged for up to an hour, while i-Log collects running applications and the time at which their running every 5 seconds (on average). To obtain a more realistic understanding of usage, screen status information was added to filter out application logs recorded while students were not actually interacting with their phone. This operation results in a dataset of 135322 applications logging events covering the 7 days of the experiment during which the time diaries were administered.
    \item \textbf{Academic performance:} Information about GPA and CFU is provided from the University of Trento and it concerns the final performance of students at the end of their first academic year (September 2016).
\end{itemize}

Notice that the overall sample considered is 67 students and not 72 due to unexpected incompatibility of some students' smartphone operating system in providing the correct logs for running applications and for one outlier case in terms of number of CFU, most likely due to incorrect recording of that information from the administration, which made it impossible to understand his total amount of credits. Even though this student has logs of his apps usage, our analysis ignore these data.


%% file: results.tex
\section{Quantifying social media usage}
\label{sec:usage}

We quantify two dimensions of the parameters for understanding the usage behaviour of students with respect to social media, i.e., $\bar S$, $\bar D$, and $\bar I$: temporal distribution of the usage, which allows us to identify temporal patterns, in Section \ref{sec:distribution}, and 
its average mean in Section \ref{sec:mean}.

\subsection{Distribution}
\label{sec:distribution}
\begin{figure}
\centering
\caption{Usage parameters of social media apps (SM) while studying}
\begin{subfigure}{.5\textwidth}
  \centering
  \caption{Sessions}
  \includegraphics[width=1.0\linewidth]{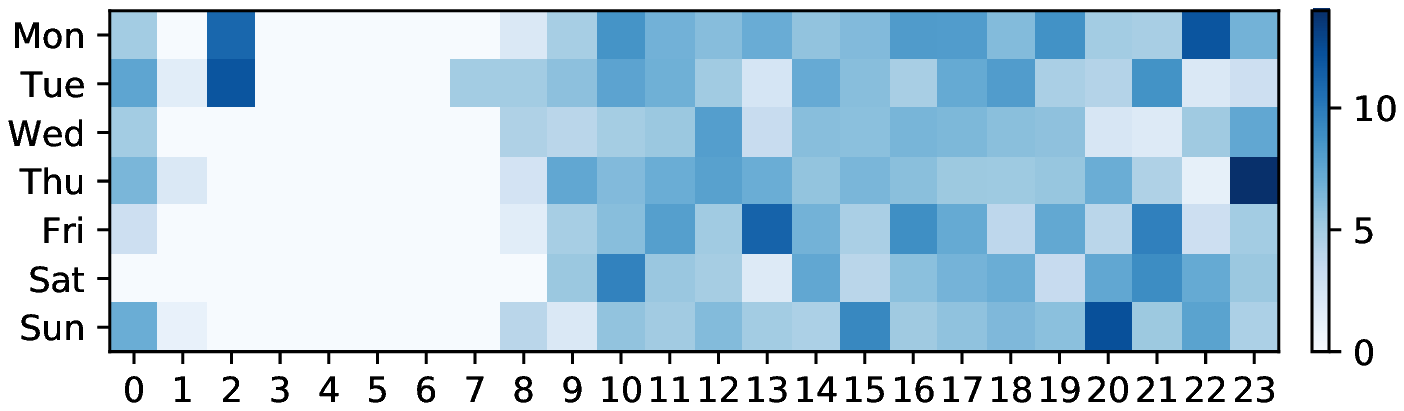}
  \label{fig:studysocsess}
\end{subfigure}%
\begin{subfigure}{.5\textwidth}
  \centering
 \caption{Duration}
  \includegraphics[width=1.0\linewidth]{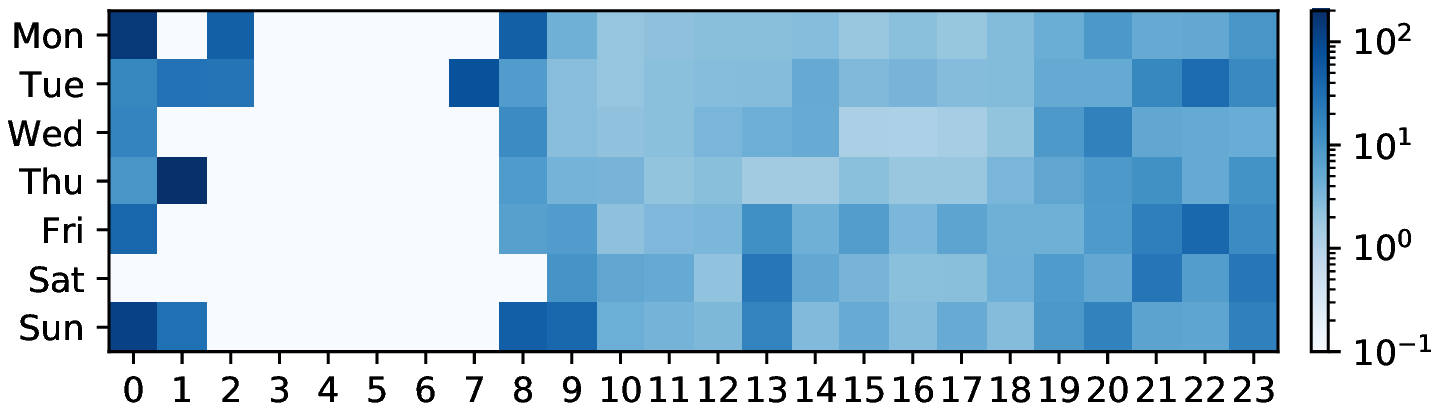}
  \label{fig:studysocdur}
\end{subfigure}%
\label{fig:distributionstudy}
\end{figure}

\begin{figure}
\centering
\caption{Usage parameters of social media apps (SM) while attending lessons}
\begin{subfigure}{.5\textwidth}
  \centering
  \caption{Sessions}
  \includegraphics[width=1.0\linewidth]{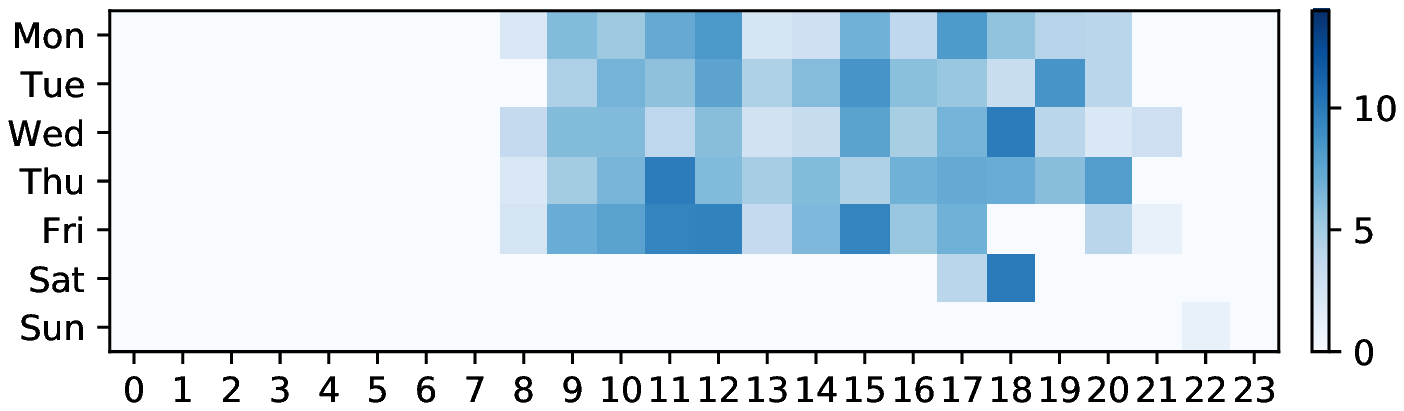}
  \label{fig:lessonsocsess}
\end{subfigure}%
\begin{subfigure}{.5\textwidth}
  \centering
  \caption{Duration}
  \includegraphics[width=1.0\linewidth]{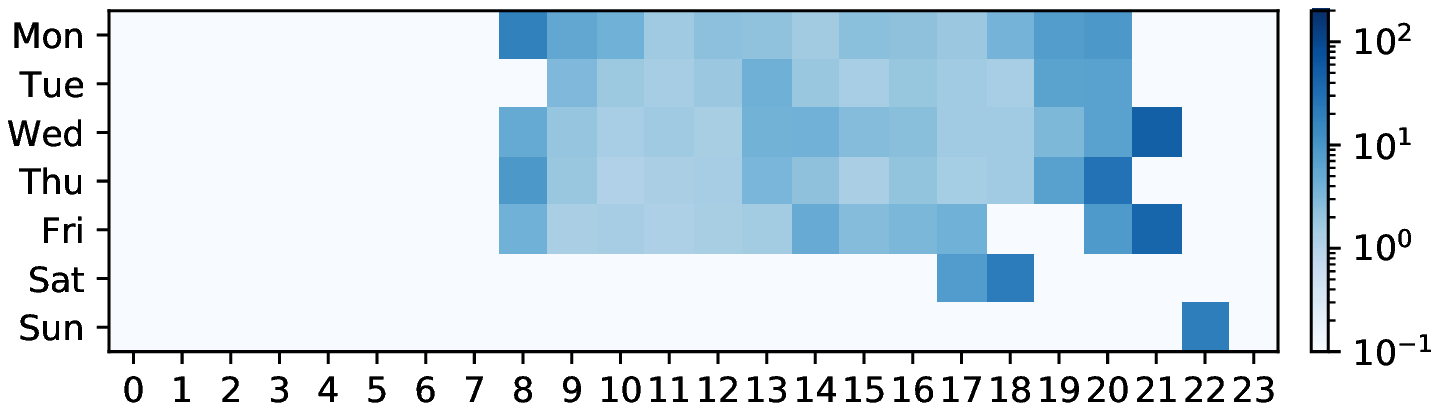}
  \label{fig:lessonsocdur}
\end{subfigure}%
\label{fig:distributionlesson}
\end{figure}

Figure \ref{fig:studysocsess} and Figure \ref{fig:studysocdur} report the distribution of $\bar S$ and $\bar D$ of social media usage during study, while Figure \ref{fig:lessonsocsess} and Figure \ref{fig:lessonsocdur} illustrate the distribution of the same parameters for attending lessons. These figures show 1 hour timeslots for every hour of the day and every day of the week, on the x and y axis respectively. Slots of empty spaces (in white) mean that no students was either studying or attending classes on that day during the specific time interval, while the darker the shade of colour the higher the number of students using apps at that time. Notice that we do not show $\bar I$ because the distribution was entirely random.

Figure \ref{fig:studysocsess} shows $\bar S$ while students reported that they were studying. Notice that the empty spaces concentrate during nighttime, as expected. Apart from few slots where there is a noticeable increase of usage, distribution through days and hours is mostly uniform. This means that students use social media while studying regardless of the time of the episode. Figure \ref{fig:studysocdur} represents the $\bar D$ of all social media apps use during study period. This parameter increases in two different portions of the day: \textit{i)} during nighttime, especially between midnight and 3AM, and \textit{ii)} early in the morning between 7AM and 9AM. Conversely, a decrease of $\bar D$ is underlined by the lighter area between 9AM and 6PM during  weekdays.

\par In the University of Trento, lessons are concentrated during working days from 8AM to 8PM. Instance of classes during the weekend and outside this range are most likely outliers due to mistakes in reporting or possibly classes of non-academic subjects. Figure \ref{fig:lessonsocsess} shows that, similarly to Figure \ref{fig:studysocsess}, the use of social media app ignores any actual time interval, although Friday appears to have a higher number of students checking SM apps. Figure \ref{fig:lessonsocdur} shows the $\bar D$ of checking social media apps during lessons. The pattern that emerges suggests that students tend to stay connected longer early in the morning and then returning to similar $\bar D$ levels as the day goes by, especially during Friday afternoon. 
Overall, in terms of temporal distribution, all parameters tend to be uniformly distributed during the whole week, be they during studying or while attending lessons. This could be interpreted as an instance of the multitasking \citep{lepp2015relationship} behaviour permeating the everyday life of students. 

\subsection{Average mean}
\label{sec:mean}

We  propose an analysis of the mean values of  $\bar S$, $\bar D$, and $\bar I$ for all apps, focusing on SM apps, with respect to activities in general, Table \ref{tab:gparameters}, while students were studying, Table \ref{tab:sparameters}, and attending lessons, Table \ref{tab:lparameters}. For each table, rows represent the type of activity (general, studying or attending lessons), the type of apps considered (all apps, SM apps as a whole, SNS, IM, and Web) and their respective parameters, while  columns represent the parameters mean usage values (Mean), their standard deviation (Sd) and the number of students (N).

\begin{table}[htbp!]
\centering
\renewcommand*{\arraystretch}{1.1}
\caption{All social media usage with respect to our variables (parameters and apps) during:}
\label{tab:gparameters}
\begin{subtable}{.3\linewidth}
\caption{General activities}
\scalebox{0.55}{
\begin{tabular}{Llllll}
\cline{4-6}
&&& \textbf{Mean}    & \textbf{Sd}      & \textbf{N} \\
       \hline\hline
\multirow{15}{*}{\textbf{\rot{General}}}  & \multirow{3}{*}{\textbf{All}}&$\mathbf{\bar S}$& 1975,55 & 798,31  & 67  \\
       &              &$ \mathbf{\bar D} $& 47,67   & 50,79   & 67  \\
       &              &$ \mathbf{\bar I} $& 236,37  & 136   & 67  \\\cline{2-6}
       & \multirow{3}{*}{\textbf{SM}} & $\mathbf{\bar S}$ & 664,25  & 360,50  & 67 \\
       &              &$ \mathbf{\bar D} $& 69,13   & 22,65    & 67  \\
       &              &$ \mathbf{\bar I} $& 157,80  & 143,65  & 67  \\\cline{2-6}
       & \multirow{3}{*}{\textbf{SNS}}& $\mathbf{\bar S}$ & 160,91  & 149,28  & 67  \\
       &              &$ \mathbf{\bar D} $& 140,25  & 96,28    & 66  \\
       &              &$ \mathbf{\bar I} $& 79,57 & 103,21  & 66  \\\cline{2-6}
       & \multirow{3}{*}{\textbf{IM}}           & $\mathbf{\bar S}$ & 440  & 282,58  & 67 \\
       &              &$ \mathbf{\bar D} $& 43,77   & 20,24   & 67  \\
       &              &$ \mathbf{\bar I} $& 180,89 & 155,24  & 67  \\\cline{2-6}
       & \multirow{3}{*}{\textbf{Web}} &     $\mathbf{\bar S}$ & 63,43   & 64,18   & 67 \\
       &              &$ \mathbf{\bar D} $& 98,71   & 40,46   & 60  \\
       &              &$ \mathbf{\bar I} $& 57,21 & 68,92 & 60  \\
       \hline
\end{tabular}
}
\end{subtable}%
    \begin{subtable}{.3\linewidth}
      \centering
        \caption{Studying}
        \label{tab:sparameters}
        \scalebox{0.55}{
        \begin{tabular}{Lllllll}
\cline{4-6}
    &&& \textbf{Mean}    & \textbf{Sd}      & \textbf{N} \\
       \hline\hline
       \multirow{15}{*}{\textbf{\rot{Study}}}  & \multirow{3}{*}{\textbf{All}}  &        $\mathbf{\bar S}$ & 296,37  & 228,99   & 67  \\
       &              &$ \mathbf{\bar D} $& 44,52   & 20,88   & 64  \\
       &              &$ \mathbf{\bar I} $& 198,64  & 148,06  & 64  \\\cline{2-6}
       & \multirow{3}{*}{\textbf{SM}} & $\mathbf{\bar S}$ & 108,44 & 96,94   & 67  \\
       &              &$ \mathbf{\bar D} $& 70,57  & 34,13   & 64  \\
       &              &$ \mathbf{\bar I} $& 121,69  & 97,38   & 64  \\\cline{2-6}
       & \multirow{3}{*}{\textbf{SNS}} & $\mathbf{\bar S}$ & 23,91   & 32,21   & 67  \\
       &              &$ \mathbf{\bar D} $& 121,37  & 100,29   & 57  \\
       &              &$ \mathbf{\bar I} $& 94,99 & 224,91 & 57  \\\cline{2-6}
       & \multirow{3}{*}{\textbf{IM}} & $\mathbf{\bar S}$ & 73,74   & 70,52 & 67  \\
       &              &$ \mathbf{\bar D} $& 49,86   & 30,57   & 64  \\
       &              &$ \mathbf{\bar I} $& 140,30 & 127,88  & 64  \\\cline{2-6}
       & \multirow{3}{*}{\textbf{Web}}  & $\mathbf{\bar S}$ & 10,79   & 16,06   & 67 \\
       &              &$ \mathbf{\bar D} $& 93,04   & 60,98   & 49  \\
       &              &$ \mathbf{\bar I} $& 58,57 & 104,83 & 49  \\
       \hline
        \end{tabular}}
    \end{subtable}%
    \begin{subtable}{.3\linewidth}
        \caption{Attending lessons}
        \label{tab:lparameters}
        \scalebox{0.55}{
        \begin{tabular}{Llllll}
\cline{4-6}
&&& \textbf{Mean}    & \textbf{Sd}      & \textbf{N}  \\
\hline\hline
       \multirow{15}{*}{\textbf{\rot{Lesson}}} & \multirow{3}{*}{\textbf{All}}          & $\mathbf{\bar S}$ & 269,97  & 176,64  & 67  \\
       &              &$ \mathbf{\bar D} $& 36,23   & 17,29   & 66   \\
       &              &$ \mathbf{\bar I} $& 167,07  & 122,23  & 66   \\\cline{2-6}
       & \multirow{3}{*}{\textbf{SM}} & $\mathbf{\bar S}$ & 87,71   & 67,37   & 67   \\
       &              &$ \mathbf{\bar D} $& 57,03   & 28,82   & 65   \\
       &              &$ \mathbf{\bar I} $& 134,21  & 203,93   & 65   \\\cline{2-6}
       & \multirow{3}{*}{\textbf{SNS}}          & $\mathbf{\bar S}$ & 19,76   & 26,28   & 67 \\
       &              &$ \mathbf{\bar D} $& 117,50  & 111,11  & 57   \\
       &              &$ \mathbf{\bar I} $& 66,65 & 87,27  & 57   \\\cline{2-6}
       & \multirow{3}{*}{\textbf{IM}}           & $\mathbf{\bar S}$ & 57,22   & 48,98   & 67   \\
       &              &$ \mathbf{\bar D} $& 36,65    & 25,74   & 65  \\
       &              &$ \mathbf{\bar I} $& 144,80  & 200,07  & 65  \\\cline{2-6}
       & \multirow{3}{*}{\textbf{Web}}      & $\mathbf{\bar S}$ & 10,73   & 14,31   & 67  \\
       &              &$ \mathbf{\bar D} $& 90,38   & 73,59   & 51  \\
       &              &$ \mathbf{\bar I} $& 87,82  & 280,39 & 51  \\
       \hline
        \end{tabular}}
    \end{subtable} 
\end{table}

\begin{sidewaystable}[ph!]
\centering
\renewcommand*{\arraystretch}{1.2}
\caption{Correlations of all apps and social media apps, with academic performance, based on overall activities plus studying and attending lessons}
\label{tab:correlation}
\scalebox{0.5}{
\begin{tabular}{|l|c|r|r|r|r|r|r|r|r|r|r|r|r|r|r|r|r|r|r|r|}
\cline{4-21}
\multicolumn{2}{c}{}&&\multicolumn{9}{c|}{\textbf{ECTS}} &\multicolumn{9}{c|}{\textbf{GPA}} \\\cline{4-21}
\multicolumn{2}{c}{}&& \textbf{All} &\textbf{Hum.}\tnote{} & \textbf{Sci.} & \textbf{F}  & \textbf{M} & \textbf{Sci.\textbackslash F} & \textbf{Hum.\textbackslash F}   & \textbf{Sci.\textbackslash M}& \textbf{Hum.\textbackslash M}& \textbf{All}   & \textbf{Hum.} & \textbf{Sci.}  & \textbf{F}  & \textbf{M}    & \textbf{Sci.\textbackslash F} & \textbf{Hum.\textbackslash F }& \textbf{Sci.\textbackslash M}& \textbf{Hum.\textbackslash M }\\
\hline
\multirow{15}{*}{\textbf{\rot{General}}} & & $\mathbf{\bar S}$  & -0.09 (68)    & 0.01 (30)      & -0.2 (38)      & 0.05 (28)     & -0.2 (40)      & 0.33 (10)    & 0.11 (16)      & -0.41 (21)     & -0.05 (14)    & -0.05 (68)    & -0.2 (30)      & 0.02 (38)      & -0.13 (28)    & -0.0 (40)      & 0.29 (10)    & -0.31 (16)     & -0.24 (21)     & -0.0 (14)\\\cline{3-21}
& & $\mathbf{\bar D}$ & 0.01 (68)     & -0.04 (30)   & -0.0 (38)    & -0.01 (28)  & -0.02 (40)   & -0.25 (10) & -0.11 (16)   & 0.05 (21)    & -0.15 (14)  & 0.07 (68)     & 0.07 (30)    & -0.06 (38)   & 0.08 (28)   & -0.03 (40)   & -0.4 (10)  & 0.05 (16)    & 0.06 (21)    & -0.11 (14)    \\\cline{3-21}
& \multirow{-3}{*}{\small{\textbf{\rot{All}}}} & $\mathbf{\bar I}$ & 0.17 (67)                     & -0.02 (29)         &\cellcolor{red!30} 0.36* (38)          & 0.1 (27)      & 0.27 (40)    & 0.13 (10)                & -0.07 (15)               & 0.41 (21)              & -0.02 (14)             & 0.16 (67)    & 0.14 (29)          &\cellcolor{red!30} 0.19* (38)          & 0.14 (27)     & 0.21 (40)    & -0.69 (10)               & 0.2 (15)                 & 0.48 (21)              & -0.04 (14)\\\cline{2-21}
& \multirow{3}{*}{\small{\textbf{\rot{SM}}}} & $\mathbf{\bar S}$   & -0.16 (67)                    & -0.02 (29)         & -0.26 (38)          & -0.01 (27)    & -0.25 (40)   & 0.05 (10)                & 0.04 (15)                &\cellcolor{red!30} -0.49* (21)            & -0.02 (14)             & -0.1 (67)    & -0.32 (29)         & 0.0 (38)            & -0.07 (27)    & -0.1 (40)    & 0.23 (10)                & -0.33 (15)               &\cellcolor{red!30} -0.3* (21)             & -0.25 (14) \\\cline{3-21}
&& $\mathbf{\bar D}$   & -0.04 (67)                    & -0.23 (29)         & 0.07 (38)           & -0.26 (27)    & 0.1 (40)     & -0.13 (10)               & -0.36 (15)               & 0.15 (21)              & -0.05 (14)             & 0.03 (67)    & 0.11 (29)          & 0.0 (38)            & -0.28 (27)    & 0.19 (40)    & -0.7 (10)                & -0.02 (15)               & 0.19 (21)              & 0.31 (14)     \\\cline{3-21}
&& $\mathbf{\bar I}$  & 0.06 (67)                     & 0.01 (29)          & 0.09 (38)           & 0.17 (27)     & -0.02 (40)   & 0.19 (10)                & 0.02 (15)                & 0.04 (21)              & -0.2 (14)              & 0.07 (67)    & 0.08 (29)          & 0.05 (38)           & 0.14 (27)     & 0.04 (40)    & -0.37 (10)               & 0.1 (15)                 & 0.15 (21)              & -0.09 (14) \\\cline{2-21}
& \multirow{3}{*}{\rot{\small{\textbf{SNS}}}}& $\mathbf{\bar S}$  & \cellcolor{red!30}-0.26* (67)                   & -0.1 (29)          &\cellcolor{red!60} \cellcolor{red!60}-0.46** (38)        & -0.03 (27)    & \cellcolor{red!60}-0.41** (40) & -0.29 (10)               & 0.08 (15)                &\cellcolor{red!30}-0.5* (21)             & -0.26 (14)             & -0.19 (67)   & -0.18 (29)         & \cellcolor{red!60}-0.31** (38)        & -0.26 (27)    & \cellcolor{red!60}-0.16** (40) & -0.23 (10)               & -0.37 (15)               &\cellcolor{red!30} -0.32* (21)            & 0.12 (14)     \\\cline{3-21}
 &  &$\mathbf{\bar D}$& 0.01 (66)                     & -0.29 (28)         & 0.2 (38)            & -0.08 (26)    & 0.05 (40)    & 0.06 (10)                & -0.42 (14)               & 0.2 (21)               & -0.24 (14)             & 0.05 (66)    & -0.02 (28)         & 0.06 (38)           & -0.03 (26)    & 0.09 (40)    & -0.64 (10)               & 0.08 (14)                & 0.08 (21)              & -0.03 (14)\\\cline{3-21}
&  & $\mathbf{\bar I}$  & -0.18 (66)                    & -0.16 (28)         & -0.19 (38)          & 0.02 (26)     & -0.26 (40)   & -0.42 (10)               & -0.26 (14)               & -0.28 (21)             & -0.09 (14)             & -0.16 (66)   & -0.11 (28)         & -0.19 (38)          & -0.06 (26)    & -0.17 (40)   & -0.33 (10)               & -0.22 (14)               & -0.2 (21)              & -0.02 (14)\\\cline{2-21}
& \multirow{3}{*}{\rot{\small{\textbf{IM}}}}             & $\mathbf{\bar S}$  & -0.1 (67)                     & 0.03 (29)          & -0.16 (38)          & -0.01 (27)    & -0.16 (40)   & 0.11 (10)                & 0.03 (15)                & -0.4 (21)              & 0.08 (14)              & -0.07 (67)   & -0.29 (29)         & 0.06 (38)           & 0.03 (27)     & -0.11 (40)   & 0.27 (10)                & -0.17 (15)               & -0.27 (21)             & -0.38 (14)\\\cline{3-21}
&  & $\mathbf{\bar D}$   & -0.06 (67)                    & -0.01 (29)         & -0.05 (38)          & -0.21 (27)    & -0.02 (40)   & 0.06 (10)                & \cellcolor{red!60}-0.75** (15)             & -0.08 (21)             & 0.38 (14)              & -0.04 (67)   & 0.08 (29)          & -0.04 (38)          & -0.32 (27)    & 0.03 (40)    & -0.34 (10)               & \cellcolor{red!60}-0.3** (15)              & 0.04 (21)              & 0.25 (14)   \\\cline{3-21}
&  & $\mathbf{\bar I}$   & 0.12 (67)                     & 0.06 (29)          & 0.15 (38)           & 0.22 (27)     & 0.06 (40)    & 0.22 (10)                & 0.1 (15)                 & 0.12 (21)              & -0.11 (14)             & 0.12 (67)    & 0.12 (29)          & 0.11 (38)           & 0.18 (27)     & 0.1 (40)     & -0.34 (10)               & 0.19 (15)                & 0.23 (21)              & -0.05 (14)\\\cline{2-21}
& \multirow{3}{*}{\rot{\small{\textbf{Web}}}}        & $\mathbf{\bar S}$  & 0.13 (67)                     & 0.01 (29)          & 0.2 (38)            & 0.01 (27)     & 0.2 (40)     & 0.14 (10)                & -0.11 (15)               & 0.22 (21)              & 0.1 (14)               & 0.19 (67)    & -0.06 (29)         & 0.38 (38)           & 0.08 (27)     & 0.26 (40)    & 0.26 (10)                & -0.09 (15)               & 0.28 (21)              & -0.07 (14)  \\\cline{3-21}
&  & $\mathbf{\bar D}$   & -0.03 (60)& \cellcolor{red!60}-0.59** (26)       & 0.21 (34)           & -0.15 (24)    & 0.05 (36)    & 0.26 (9)                 & -0.5 (13)                & 0.23 (18)              &\cellcolor{red!30} \cellcolor{red!30}-0.68* (13)            & -0.0 (60)    & \cellcolor{red!60}-0.15** (26)       & 0.09 (34)           & -0.23 (24)    & 0.14 (36)    & -0.19 (9)                & -0.23 (13)               & 0.08 (18)              &\cellcolor{red!30} -0.0* (13)\\\cline{3-21}
&  & $\mathbf{\bar I}$ & 0.12 (60) & -0.02 (26)         & 0.2 (34)            & -0.0 (24)     & 0.18 (36)    & -0.19 (9)                & 0.05 (13)                & 0.3 (18)               & -0.22 (13)             & 0.2 (60)     & 0.19 (26)          & 0.25 (34)           & 0.31 (24)     & 0.17 (36)    & 0.06 (9)                 & 0.44 (13)                & 0.38 (18)              & -0.18 (13)\\\hline

\multirow{15}{*}{\textbf{\rot{Study}}}  & \multirow{3}{*}{\rot{\small{\textbf{All}}}} &$\mathbf{\bar S}$&  0.02 (67) & 0.11 (29)          & -0.17 (38)          & 0.14 (27)     & -0.06 (40)   & 0.34 (10)                & -0.04 (15)               & -0.24 (21)             & 0.23 (14)              & -0.04 (67)   & -0.12 (29)         & -0.12 (38)          & -0.09 (27)    & -0.04 (40)   & 0.17 (10)                & -0.45 (15)               & -0.11 (21)             & 0.06 (14)\\\cline{3-21}
&  & $\mathbf{\bar D}$  & -0.12 (65)                    & -0.33 (28)         & -0.09 (37)          & -0.06 (27)    & -0.2 (38)    & -0.05 (10)               & -0.12 (15)               & 0.01 (20)              &\cellcolor{red!30} -0.67* (13)            & 0.02 (65)    & 0.01 (28)          & -0.08 (37)          & 0.15 (27)     & -0.1 (38)    & -0.17 (10)               & 0.26 (15)                & 0.04 (20)              &\cellcolor{red!30} -0.29* (13)\\\cline{3-21}
&  & $\mathbf{\bar I}$ & 0.23 (65)                     & -0.02 (28)         &\cellcolor{red!30} 0.4* (37)           & 0.11 (27)     & \cellcolor{red!30}0.36* (38)   & -0.15 (10)               & -0.02 (15)               &\cellcolor{red!30} 0.52* (20)             & -0.16 (13)             & \cellcolor{red!30}0.26* (65)   & 0.29 (28)          &\cellcolor{red!30} 0.28* (37)          & 0.15 (27)     &\cellcolor{red!30} 0.37* (38)   & -0.51 (10)               & 0.33 (15)                &\cellcolor{red!30} 0.57* (20)             & 0.21 (13)\\\cline{2-21}
 & \multirow{3}{*}{\rot{\small{\textbf{SM}}}}  &$\mathbf{\bar S}$ & -0.03 (67) & 0.16 (29)          & -0.26 (38)          & 0.07 (27)     & -0.11 (40)   & -0.07 (10)               & 0.12 (15)                & -0.31 (21)             & 0.21 (14)              & -0.08 (67)   & -0.16 (29)         & -0.14 (38)          & -0.05 (27)    & -0.11 (40)   & 0.02 (10)                & -0.33 (15)               & -0.18 (21)             & -0.06 (14)\\\cline{3-21}
 && $\mathbf{\bar D}$ & -0.07 (64) &\cellcolor{red!30} -0.41* (28)        & 0.02 (36)           & -0.29 (26)    & 0.08 (38)    & -0.19 (10)               &\cellcolor{red!30} -0.62* (15)              & 0.17 (20)              & -0.14 (13)             & -0.04 (64)   &\cellcolor{red!30} \cellcolor{red!30} -0.14* (28)        & -0.06 (36)          & -0.24 (26)    & 0.06 (38)    & -0.4 (10)                &\cellcolor{red!30} -0.12* (15)              & 0.17 (20)              & -0.08 (13)\\\cline{3-21}
&  & $\mathbf{\bar I}$ & 0.14 (64) & -0.17 (28)         & 0.3 (36)            & 0.24 (26)     & 0.15 (38)    & 0.27 (10)                & 0.08 (15)                & 0.33 (20)              & -0.34 (13)             & 0.11 (64)    & 0.13 (28)          & 0.18 (36)           & 0.06 (26)     & 0.2 (38)     & -0.22 (10)               & 0.25 (15)                & 0.44 (20)              & 0.12 (13) \\\cline{2-21}
& \multirow{3}{*}{\rot{\small{\textbf{SNS}}}} & $\mathbf{\bar S}$ & -0.11 (67) & 0.02 (29)          & \cellcolor{red!60}-0.49** (38)        & 0.08 (27)     & -0.22 (40)   & -0.46 (10)               & 0.09 (15)                &\cellcolor{red!30} -0.48* (21)            & -0.03 (14)             & -0.15 (67)   & -0.17 (29)         & \cellcolor{red!60}-0.41** (38)        & -0.22 (27)    & -0.12 (40)   & -0.54 (10)               & -0.39 (15)               &\cellcolor{red!30} \cellcolor{red!30} -0.33* (21)            & 0.01 (14)   \\\cline{3-21}
&  & $\mathbf{\bar D}$  & 0.17 (57)                     & -0.01 (25)         & 0.23 (32)           & -0.09 (23)    & 0.29 (34)    & 0.08 (9)                 & -0.38 (13)               & 0.34 (18)              & 0.28 (12)              & 0.02 (57)    & -0.13 (25)         & 0.07 (32)           & -0.18 (23)    & 0.12 (34)    & -0.36 (9)                & -0.08 (13)               & 0.23 (18)              & -0.06 (12)  \\\cline{3-21}
&  & $\mathbf{\bar I}$   & -0.01 (57)                    & -0.03 (25)         & 0.11 (32)           & 0.03 (23)     & 0.01 (34)    & 0.25 (9)                 & 0.32 (13)                & 0.09 (18)              & -0.25 (12)             & 0.0 (57)     & -0.12 (25)         & 0.13 (32)           & -0.08 (23)    & 0.07 (34)    & -0.11 (9)                & 0.18 (13)                & 0.14 (18)              & -0.24 (12)\\\cline{2-21}
& \multirow{3}{*}{\rot{\small{\textbf{IM}}}}& $\mathbf{\bar S}$ & -0.01 (67)                    & 0.22 (29)          & -0.19 (38)          & 0.04 (27)     & -0.06 (40)   & -0.0 (10)                & 0.12 (15)                & -0.27 (21)             & 0.32 (14)              & -0.05 (67)   & -0.12 (29)         & -0.07 (38)          & -0.0 (27)     & -0.1 (40)    & 0.09 (10)                & -0.21 (15)               & -0.17 (21)             & -0.06 (14)\\\cline{3-21}
&  &$\mathbf{\bar D}$ & -0.06 (64)                    & -0.29 (28)         & 0.02 (36)           & -0.22 (26)    & 0.04 (38)    & -0.0 (10)                & \cellcolor{red!80}-0.79*** (15)            & -0.03 (20)             & 0.22 (13)              & -0.08 (64)   & -0.08 (28)         & -0.08 (36)          & -0.3 (26)     & 0.01 (38)    & -0.33 (10)               & \cellcolor{red!80}-0.2*** (15)             & 0.02 (20)              & 0.01 (13)\\\cline{3-21}
&  & $\mathbf{\bar I}$  &\cellcolor{red!30} 0.29* (64)                    & -0.12 (28)         &\cellcolor{red!60} 0.46** (36)         & 0.29 (26)     & \cellcolor{red!30}0.33* (38)   & 0.22 (10)                & 0.26 (15)                &\cellcolor{red!60} 0.56** (20)            & -0.33 (13)             & 0.23 (64)    & 0.17 (28)          &\cellcolor{red!60} 0.3** (36)          & 0.06 (26)     &\cellcolor{red!30} 0.35* (38)   & -0.18 (10)               & 0.24 (15)                &\cellcolor{red!60} 0.61** (20)            & 0.23 (13)\\\cline{2-21}
& \multirow{3}{*}{\rot{\small{\textbf{Web}}}}& $\mathbf{\bar S}$ & 0.07 (67)                     & 0.07 (29)          & 0.03 (38)           & 0.08 (27)     & 0.07 (40)    & -0.14 (10)               & -0.05 (15)               & 0.13 (21)              & 0.12 (14)              & 0.03 (67)    & -0.07 (29)         & 0.06 (38)           & 0.09 (27)     & -0.01 (40)   & -0.05 (10)               & -0.05 (15)               & 0.19 (21)              & -0.17 (14)\\\cline{3-21}
&  & $\mathbf{\bar D}$& -0.18 (49)                    &\cellcolor{red!30} -0.43* (23)        & -0.15 (26)          & -0.3 (20)     & -0.1 (29)    & -0.07 (7)                & -0.54 (13)               & -0.12 (16)             & -0.26 (10)             & 0.09 (49)    &\cellcolor{red!30} -0.07* (23)        & 0.13 (26)           & 0.12 (20)     & 0.07 (29)    & 0.43 (7)                 & -0.07 (13)               & 0.1 (16)               & 0.1 (10)\\\cline{3-21}
&  & $\mathbf{\bar I}$ & 0.05 (49)                     & -0.17 (23)         & 0.19 (26)           & -0.13 (20)    & 0.16 (29)    & -0.5 (7)                 & -0.01 (13)               & 0.39 (16)              & -0.39 (10)             & 0.08 (49)    & 0.21 (23)          & -0.02 (26)          & 0.15 (20)     & 0.06 (29)    & -0.49 (7)                & 0.43 (13)                & 0.1 (16)               & -0.02 (10)\\\hline
\multirow{15}{*}{\rot{\textbf{Lesson}}} & \multirow{3}{*}{\rot{\small{\textbf{All}}}} & $\mathbf{\bar S}$ & 0.02 (67)                     & -0.06 (29)         & 0.1 (38)            & -0.05 (27)    & 0.07 (40)    & 0.4 (10)                 & -0.18 (15)               & -0.11 (21)             & 0.09 (14)              & 0.03 (67)    & -0.26 (29)         & 0.24 (38)           & -0.25 (27)    & 0.17 (40)    & 0.36 (10)                & -0.62 (15)               & -0.07 (21)             & 0.06 (14)\\\cline{3-21}
&  &$\mathbf{\bar D}$& -0.11 (66)                    & -0.37 (29)         & 0.05 (37)           & -0.14 (27)    & -0.08 (39)   & -0.01 (10)               & -0.5 (15)                & -0.01 (20)             & -0.22 (14)             & -0.08 (66)   & -0.17 (29)         & 0.06 (37)           & -0.25 (27)    & 0.06 (39)    & -0.2 (10)                & -0.41 (15)               & 0.16 (20)              & 0.07 (14)\\\cline{3-21}
&  & $\mathbf{\bar I}$ &\cellcolor{red!30} 0.31* (66)                    & 0.0 (29)           &\cellcolor{red!60} 0.44** (37)         & 0.33 (27)     &\cellcolor{red!30} 0.32* (39)   & 0.39 (10)                & 0.15 (15)                & \cellcolor{red!30}0.47* (20)             & -0.26 (14)             & \cellcolor{red!80}0.4*** (66)  & 0.42 (29)          & \cellcolor{red!60}0.41** (37)         & 0.42 (27)     &\cellcolor{red!30} 0.41* (39)   & -0.09 (10)               & 0.64 (15)                & \cellcolor{red!30} 0.65* (20)             & 0.14 (14)\\\cline{2-21}
& \multirow{3}{*}{\rot{\small{\textbf{SM}}}} & $\mathbf{\bar S}$  & -0.02 (67)                    & 0.12 (29)          & -0.02 (38)          & -0.01 (27)    & -0.03 (40)   & 0.19 (10)                & 0.21 (15)                & -0.24 (21)             & 0.06 (14)              & -0.01 (67)   & -0.3 (29)          & 0.17 (38)           & -0.17 (27)    & 0.07 (40)    & 0.27 (10)                & -0.49 (15)               & -0.15 (21)             & -0.14 (14)\\\cline{3-21}
&  & $\mathbf{\bar D}$ & -0.21 (65)                    & -0.26 (29)         & -0.15 (36)          & \cellcolor{red!30} -0.47* (26)   & -0.11 (39)   & -0.38 (9)                & -0.45 (15)               & -0.12 (20)             & -0.07 (14)             & -0.11 (65)   & 0.01 (29)          & -0.11 (36)          &\cellcolor{red!30} -0.51* (26)   & 0.05 (39)    & -0.56 (9)                & -0.35 (15)               & -0.01 (20)             & 0.31 (14)\\\cline{3-21}
&  & $\mathbf{\bar I}$ & 0.16 (65)                     & -0.13 (29)         & 0.24 (36)           & 0.28 (26)     & 0.16 (39)    & 0.59 (9)                 & 0.02 (15)                & 0.23 (20)              & -0.24 (14)             & 0.2 (65)     & 0.11 (29)          & 0.29 (36)           & -0.11 (26)    & 0.31 (39)    & -0.36 (9)                & 0.0 (15)                 & 0.44 (20)              & 0.22 (14)\\\cline{2-21}
& \multirow{3}{*}{\rot{\small{\textbf{SNS}}}}        &$\mathbf{\bar S}$ & -0.24 (67)                    & -0.0 (29)          & \cellcolor{red!30}-0.34* (38)         & -0.16 (27)    & -0.3 (40)    & -0.33 (10)               & 0.02 (15)                & -0.37 (21)             & -0.09 (14)             & -0.21 (67)   & -0.21 (29)         & \cellcolor{red!30} -0.2* (38)          & -0.46 (27)    & -0.08 (40)   & -0.3 (10)                & -0.55 (15)               & -0.23 (21)             & 0.43 (14)\\\cline{3-21}
&  & $\mathbf{\bar D}$  & -0.24 (57)                    &\cellcolor{red!30} -0.45* (26)        & -0.11 (31)          &\cellcolor{red!60} -0.55** (22)  & -0.08 (35)   & -0.38 (8)                & \cellcolor{red!80}-0.84*** (12)            & -0.03 (17)             & -0.18 (14)             & -0.13 (57)   &\cellcolor{red!30} -0.15* (26)        & -0.2 (31)           & \cellcolor{red!60}-0.33** (22)  & -0.01 (35)   & -0.52 (8)                & \cellcolor{red!80}-0.35*** (12)            & -0.13 (17)             & 0.01 (14)   \\\cline{3-21}
&  & $\mathbf{\bar I}$ & 0.07 (57)                     & -0.08 (26)         & 0.21 (31)           & 0.2 (22)      & -0.02 (35)   & 0.08 (8)                 & 0.21 (12)                & 0.01 (17)              & -0.21 (14)             & 0.05 (57)    & 0.09 (26)          & 0.08 (31)           & -0.13 (22)    & 0.2 (35)     & -0.07 (8)                & -0.17 (12)               & 0.1 (17)               & 0.31 (14)   \\\cline{2-21}
& \multirow{3}{*}{\rot{\textbf{\small{IM}}}} & $\mathbf{\bar S}$ & 0.08 (67)                     & 0.13 (29)          & 0.1 (38)            & 0.07 (27)     & 0.08 (40)    & 0.38 (10)                & 0.2 (15)                 & -0.1 (21)              & 0.12 (14)              & 0.05 (67)    & -0.27 (29)         & 0.25 (38)           & 0.01 (27)     & 0.07 (40)    & 0.45 (10)                & -0.26 (15)               & -0.08 (21)             & -0.28 (14)\\\cline{3-21}
&  & $\mathbf{\bar D}$  & -0.05 (65)                    & 0.02 (29)          & -0.04 (36)          & -0.2 (26)     & -0.01 (39)   & 0.04 (9)                 & -0.42 (15)               & -0.08 (20)             & 0.4 (14)               & -0.03 (65)   & 0.03 (29)          & -0.01 (36)          & -0.29 (26)    & 0.05 (39)    & -0.3 (9)                 & -0.23 (15)               & 0.02 (20)              & 0.26 (14)   \\\cline{3-21}
&  &$\mathbf{\bar I}$ & 0.14 (65)                     & -0.11 (29)         & 0.23 (36)           & 0.22 (26)     & 0.15 (39)    & 0.5 (9)                  & -0.04 (15)               & 0.21 (20)              & -0.17 (14)             & 0.18 (65)    & 0.08 (29)          & 0.27 (36)           & -0.14 (26)    & 0.29 (39)    & -0.4 (9)                 & 0.01 (15)                & 0.43 (20)              & 0.14 (14)\\\cline{2-21}
& \multirow{3}{*}{\rot{\small{\textbf{Web}}}} & $\mathbf{\bar S}$     & 0.09 (67)                     & 0.11 (29)          & 0.12 (38)           & 0.04 (27)     & 0.12 (40)    & 0.04 (10)                & 0.33 (15)                & 0.08 (21)              & -0.07 (14)             & 0.15 (67)    & -0.01 (29)         & 0.27 (38)           & 0.05 (27)     & 0.22 (40)    & 0.13 (10)                & 0.13 (15)                & 0.1 (21)               & -0.1 (14)\\\cline{3-21}
&  & $\mathbf{\bar D}$ & -0.15 (51)                    & -0.32 (22)         & -0.03 (29)          & -0.09 (21)    & -0.18 (30)   & 0.34 (7)                 & -0.5 (12)                & -0.32 (15)             & -0.28 (10)             & -0.19 (51)   & -0.37 (22)         & -0.06 (29)          & -0.34 (21)    & -0.14 (30)   & -0.38 (7)                & -0.25 (12)               & -0.13 (15)             & -0.48 (10)\\\cline{3-21}
&  & $\mathbf{\bar I}$ & 0.22 (51)                     & 0.29 (22)          & 0.27 (29)           &\cellcolor{red!30} 0.45* (21)    & 0.22 (30)    & 0.75 (7)                 & 0.4 (12)                 & 0.27 (15)              & -0.51 (10)             &\cellcolor{red!30} 0.3* (51)    & 0.51 (22)          & 0.36 (29)           &\cellcolor{red!30} 0.52* (21)    & 0.33 (30)    & 0.43 (7)                 & 0.57 (12)                & 0.52 (15)              & 0.33 (10) \\
                \hline
\multicolumn{21}{l}{\small{Notes: ${}^{*} p<.05$, ${}^{**} p<.01$, ${}^{***} p<.001$; Hum.=Humanities; Sci.=Scientific; F=Female; G=Gender;  (N)=N of students; the gradient of color represents the result significance.}}\
\end{tabular}
   
   } 
\end{sidewaystable}

For general activities, ${\bar S}$ appears to be the most relevant parameter both for all apps and SM apps (1975.55, SD 798.31 and 664.25, SD 360.50), followed by ${\bar I}$ and ${\bar D}$. Within SM apps, IM are the most checked type of apps, with ${\bar S}$ being almost 4 times the other apps (440, SD 282.58), but also the one with the highest value for ${\bar I}$ (180.89, SD 155,24), while SNS sessions last the longest (${\bar D}$ of 120.25). This general pattern is also true for reported usage of smartphones both while studying and attending lessons, although with some differences. Firstly, SM apps are checked more frequently and for longer periods of time while studying than during lessons (higher values of ${\bar S}$ and ${\bar D}$ for SM and each app type). Notice that in the case of ${\bar D}$ of SNS and Web the values are nonetheless very close, unlike IM, with 49.86, SD 30.57 for studying vs 36.65, SD 25.74 for attending lessons. Secondly, while $\bar I$ is lower during studying in terms of SM (121.69, SD 97.38), its values for IM apps are almost equal: for study the mean is 140.30 seconds (SD 127.88) and for lesson is 144.80 seconds (SD 200.07). Overall, these findings suggest the following:
\begin{itemize}
    \item On average, students check SM apps more frequently and for longer periods while studying than attending lessons (higher ${\bar S}$ and ${\bar D}$), but while in class these sessions are more done in a longer window of time (higher $\bar I$).
    \item Within SM apps for both studying and attending lessons, IM apps are the most checked but with the longest window of time in between sessions, while SNS apps are the ones with the highest duration of usage.
\end{itemize}

%% file: correlation.tex
\section{Social media usage vs GPA and CFU}
\label{sec:correlation}

Table \ref{tab:correlation} shows how $\bar S$, $\bar D$ and $\bar I$ are correlated to students' CFU and GPA by using Pearson's correlation. 
\par In Table \ref{tab:correlation} the darker the color of the cells whose parameters, considering columns and rows, obtain a significant value with respect to the correlation coefficient, the higher the value significance (i.e., higher \textit{p} value). Rows represent $\bar S$, $\bar D$ and $\bar I$ for the combination of application type and activities from the Section \ref{sec:usage}. Columns represent the sociodemographic variables considered and the academic performance indexes, GPA and CFU. The socio-demographics are gender, faculties (distinguishing between scientific and humanities), and the combination of the two, i.e., male and female students from either faculties. 
\par We expect a negative correlation in an increase of $\bar S$ and $\bar D$, since they would imply more smartphone usage and hence less time dedicated to academic activities. Conversely, we expect a higher value of  $\bar I$ to be positively correlated with academic activity, since it would indicate less time dedicated to smartphones. 

\subsection{Significant values of social media usage}

Table \ref{tab:general}, Table \ref{tab:study} and Table \ref{tab:lesson} summarize the occurrence of significant values for $\bar S$, $\bar D$, and $\bar I$. Columns indicate the amount of significant values, divided according to their \textit{p} value, and their total amount, while rows represent the type of activity (general, studying or attending lessons), the type of apps considered (all apps, SM apps as a whole, SNS, IM, and Web), their respective parameters, and their sum accounting for both the amount of values and their significance.
\par Table \ref{tab:general} shows that during general activities  $\bar S$ and $\bar D$ have a relatively close amount of significant correlations (9 and 6, respectively), while $\bar I$ has only 2. Moreover, $\bar S$ of SNS is especially significant, reaching 7 values (4 with $ p<.01$). 
\par If we look at the same vales for studying and attending lessons, Table \ref{tab:study} and Table \ref{tab:lesson} respectively, studying provides more occurrences than attending lessons (28 vs 21), but with similar occurrences of values per significance.

\begin{table}[!htb]
\centering
\renewcommand*{\arraystretch}{1.1}
    \caption{Number of significant value occurrences from our variables in:}
    \begin{subtable}{.3\linewidth}
\centering
\caption{General activities}
\label{tab:general}
\scalebox{0.55}{
\begin{tabular}{Lllllll}
\cline{4-7}
&&&\rot{$p<.05$}&\rot{$ p<.01$}& \rot{$ p<.001$}&\rot{\textbf{Total}}\\
       \hline\hline
\multirow{21}{*}{\rot{\textbf{General}}} & \multirow{4}{*}{\textbf{All}}&$\mathbf{\bar S}$& 0 & 0 & 0 & 0 \\
&&$ \mathbf{\bar D} $& 0 & 0 & 0 & 0 \\
&&$ \mathbf{\bar I} $& 2 & 0 & 0 & 2 \\
\cline{3-7}
&&\textbf{Total}& 2 &  0 & 0 &  2 \\\cline{2-7}
& \multirow{4}{*}{\textbf{SM}}&$\mathbf{\bar S}$& 2 & 0 & 0 & 2 \\
&&$ \mathbf{\bar D} $& 0 & 0 & 0 & 0 \\
&&$ \mathbf{\bar I} $& 0 & 0 & 0 & 0 \\
\cline{3-7}
&&\textbf{Total}& 2 & 0 & 0 & 2 \\
\cline{2-7}
& \multirow{4}{*}{\textbf{SNS}}&$\mathbf{\bar S}$& 3 & 4 & 0 & 7 \\
&&$ \mathbf{\bar D} $& 0 & 0 & 0 & 0 \\
&&$ \mathbf{\bar I} $& 0 & 0 & 0 & 0 \\
\cline{3-7}
&&\textbf{Total}& 3 & 4 & 0 & 7 \\
\cline{2-7}
& \multirow{4}{*}{\textbf{IM}}&$\mathbf{\bar S}$& 0 & 0 & 0 & 0 \\
&&$ \mathbf{\bar D} $& 0 & 2 & 0 & 0 \\
&&$ \mathbf{\bar I} $& 0 & 0 & 0 & 0 \\
\cline{3-7}
&&\textbf{Total}& 0 & 2 & 0 & 2 \\
\cline{2-7}
& \multirow{4}{*}{\textbf{Web}}&$\mathbf{\bar S}$& 0 & 0 & 0 & 0 \\
&&$ \mathbf{\bar D} $& 2 & 2 & 0 & 4 \\
&&$ \mathbf{\bar I} $& 0 & 0 & 0 & 0 \\
\cline{3-7}
&&\textbf{Total}& 2 & 2 & 0 & 4 \\
\cline{2-7}
\hhline{~|--|--|}
& \multirow{4}{*}{\textbf{Sum}}&$\mathbf{\bar S}$& 5 & 4 & 0 & 9 \\
&&$ \mathbf{\bar D} $& 2 & 4 & 0 & 6 \\
&&$ \mathbf{\bar I} $& 2 & 0 & 0 & 2 \\
\cline{3-7}
&&\textbf{Total}& 9 & 8 & 0 & 17\\
\hline
\end{tabular}}
\end{subtable}%
    \begin{subtable}{.3\linewidth}
      \centering
      \caption{Studying}
      \label{tab:study}
        \scalebox{0.55}{
\begin{tabular}{Lllllll}
\cline{4-7}
&&&\rot{$p<.05$}&\rot{$ p<.01$}& \rot{$ p<.001$}&\rot{\textbf{Total}}\\
       \hline\hline
\multirow{21}{*}{\rot{\textbf{Study}}}  & \multirow{4}{*}{\textbf{All}}&$\mathbf{\bar S}$& 0 & 0 & 0 & 0 \\
&&$ \mathbf{\bar D} $& 2 & 0 & 0 & 2 \\
&&$ \mathbf{\bar I} $& 7 & 0 & 0 & 7 \\
\cline{3-7}
&&\textbf{Total}& 9 & 0 & 0 & 9 \\\cline{2-7}
& \multirow{4}{*}{\textbf{SM}}&$\mathbf{\bar S}$& 0 & 0 & 0 & 0 \\
&&$ \mathbf{\bar D} $& 4 & 0 & 0 & 4 \\
&&$ \mathbf{\bar I} $& 0 & 0 & 0 & 0 \\
\cline{3-7}
&&\textbf{Total}& 4 & 0 & 0 & 4 \\
\cline{2-7}
& \multirow{4}{*}{\textbf{SNS}}&$\mathbf{\bar S}$& 2 & 2 & 0 & 4 \\
&&$ \mathbf{\bar D} $& 0 & 0 & 0 & 0 \\
&&$ \mathbf{\bar I} $& 0 & 0 & 0 & 0 \\
\cline{3-7}
&&\textbf{Total}& 2 & 2 & 0 & 4 \\
\cline{2-7}
& \multirow{4}{*}{\textbf{IM}}&$\mathbf{\bar S}$& 0 & 0 & 0 & 0 \\
&&$ \mathbf{\bar D} $& 0 & 0 & 2 & 2 \\
&&$ \mathbf{\bar I} $& 3 & 4 & 0 & 7 \\
\cline{3-7}
&&\textbf{Total}& 3 & 4 & 2 & 9 \\
\cline{2-7}
& \multirow{4}{*}{\textbf{Web}}&$\mathbf{\bar S}$& 0 & 0 & 0 & 0 \\
&&$ \mathbf{\bar D} $& 2 & 0 & 0 & 2 \\
&&$ \mathbf{\bar I} $& 0 & 0 & 0 & 0 \\
\cline{3-7}
&&\textbf{Total}& 2 & 0 & 0 & 2 \\
\cline{2-7}
& \multirow{4}{*}{\textbf{Sum}}&$\mathbf{\bar S}$& 2 & 2 & 0 & 4 \\
&&$ \mathbf{\bar D} $& 8 & 0 & 2 & 10 \\
&&$ \mathbf{\bar I} $& 10 & 4 & 0 & 14 \\
\cline{3-7}
&&\textbf{Total}& 20 & 6 & 2 & 28\\
\hline
\end{tabular}}
    \end{subtable}%
    \begin{subtable}{.3\linewidth}
      \centering
        \caption{Attending lessons}
        \label{tab:lesson}
        \scalebox{0.55}{
        \begin{tabular}{Lllllll}
\cline{4-7}
&&&\rot{$p<.05$}&\rot{$ p<.01$}& \rot{$ p<.001$}&\rot{\textbf{Total}}\\
       \hline\hline
\multirow{21}{*}{\rot{\textbf{Lesson}}}  & \multirow{4}{*}{\textbf{All}}&$\mathbf{\bar S}$& 0 & 0 & 0 & 0 \\
&&$ \mathbf{\bar D} $& 0 & 0 & 0 & 0 \\
&&$ \mathbf{\bar I} $& 5 & 2 & 1 & 8 \\
\cline{3-7}
&&\textbf{Total}& 5 & 2 & 1 & 8 \\\cline{2-7}
& \multirow{4}{*}{\textbf{SM}}&$\mathbf{\bar S}$& 0 & 0 & 0 & 0 \\
&&$ \mathbf{\bar D} $& 2 & 0 & 0 & 2 \\
&&$ \mathbf{\bar I} $& 0 & 0 & 0 & 0 \\
\cline{3-7}
&&\textbf{Total}& 2 & 0 & 0 & 2 \\
\cline{2-7}
& \multirow{4}{*}{\textbf{SNS}}&$\mathbf{\bar S}$& 2 & 0 & 0 & 2 \\
&&$ \mathbf{\bar D} $& 6 & 0 & 0 & 6 \\
&&$ \mathbf{\bar I} $& 0 & 0 & 0 & 0 \\
\cline{3-7}
&&\textbf{Total}& 8 & 0 & 0 & 8 \\
\cline{2-7}
& \multirow{4}{*}{\textbf{IM}}&$\mathbf{\bar S}$& 0 & 0 & 0 & 0 \\
&&$ \mathbf{\bar D} $& 0 & 0 & 0 & 0 \\
&&$ \mathbf{\bar I} $& 0 & 0 & 0 & 0 \\
\cline{3-7}
&&\textbf{Total}& 0 & 0 & 0 & 0 \\
\cline{2-7}
& \multirow{4}{*}{\textbf{Web}}&$\mathbf{\bar S}$& 0 & 0 & 0 & 0 \\
&&$ \mathbf{\bar D} $& 0 & 0 & 0 & 0 \\
&&$ \mathbf{\bar I} $& 3 & 0 & 0 & 3 \\
\cline{3-7}
&&\textbf{Total}& 3 & 0 & 0 & 3 \\
\cline{2-7}
\cline{2-7}
& \multirow{4}{*}{\textbf{Sum}}&$\mathbf{\bar S}$& 2 & 0 & 0 & 2 \\
&&$ \mathbf{\bar D} $& 8 & 0 & 0 & 8 \\
&&$ \mathbf{\bar I} $& 8 & 2 & 1 & 11 \\
\cline{3-7}
&&\textbf{Total}& 18 & 2 & 1 & 21\\
\hline
\end{tabular}}
    \end{subtable} 
\end{table}

Moreover, $\bar I$ is the parameter with the most occurrences of significant values for both activities, with a total of 25, followed by $\bar D$ with 18 and finally $\bar S$, only 6. Within SM, IM provides the most significant values for studying, concentrated on $\bar I$; however, there are no IM values for lessons, which means that IM provide no correlations in this case. On the other hand, SNS provide the most values for establishing correlations in lessons, especially for $\bar D$. 
\par Overall, these findings suggest that our parameters for usage and social media plus the time diary answers for academic activities allow us to effectively underline different patterns of SM app influence. Moreover:

\begin{itemize}
\item While studying, the average duration of usage of IM apps ($\bar D$ with negative \textit{p} values) is the most harmful for academic performance; however, the longer students avoid them ($\bar I$ with positive \textit{p} values) the higher their performances.
\item The average duration of usage ($\bar D$ with negative \textit{p} values) and the average occurrences of checking ($\bar S$ with negative \textit{p} values) SNS while attending lessons negatively affect students' academic performance
\end{itemize}

\subsection{Significant values for CFU and GPA}

Table \ref{tab:ects} and Table \ref{tab:gpa} show the total occurrences of significant values between our variables and the CFU and GPA, i.e., 33 for both. Columns indicate the type of variable considered: all, faculty (humanities and scientific), gender (females and males), the combination of the two (females and males in scientific and humanities faculties) and their sum. Rows represent the amount of significant values, divided according to their \textit{p} value, and their total amount. 
\par On average, the influence of SM apps both on CFU and GPA appears to be stronger for scientific students than for students from humanities (7 vs 4), while gender differences seem to be less important, being almost equally distributed in our sample. In addition, distinguishing within each faculty suggests that being either a male student enrolled in a scientific faculty or being a female from humanities are the most ``at risk" groups of a decrease of academic performance.

\begin{table}[htb!]
\centering
\caption{Number of significant correlations for CFU}
\centering
\label{tab:ects}
\begin{tabular}{lllllllllll}
&\multicolumn{10}{c}{\textbf{CFU}}\\
\cline{2-11}
&\rot{\textbf{All}}&\rot{\textbf{Hum.}}&\rot{\textbf{Sci.}}&\rot{\textbf{F}}&\rot{\textbf{M}}&\rot{\textbf{Sci.\textbackslash F}}&\rot{\textbf{Hum.\textbackslash F}}&\rot{\textbf{Sci.\textbackslash M}}&\rot{\textbf{Hum.\textbackslash M}}&\rot{\textbf{Sum}}\\
\hline
$p<.05$& 3 & 3 & 2 & 2 & 3 & 0 & 1 & 5 & 2 & 21\\
\hline
$p<.01$& 0 & 1 & 5 & 1 & 1 & 0 & 1 & 1 & 0 & 10\\
\hline
$ p<.001$& 0 & 0 & 0 & 0 & 0 & 0 & 2 & 0 & 0 & 2\\
\hline
\textbf{Tot}& 3 & 4 & 7 & 3 & 4 & 0 & 4 & 6 & 2 & 33\\
\hline
\end{tabular}
\end{table}%
\begin{table}[htb!]
\centering
\caption{Number of significant correlations for GPA}
\label{tab:gpa}
\begin{tabular}{lllllllllll}
&\multicolumn{10}{c}{\textbf{GPA}}\\
\cline{2-11}
&\rot{\textbf{All}}&\rot{\textbf{Hum.}}&\rot{\textbf{Sci.}}&\rot{\textbf{F}}&\rot{\textbf{M}}&\rot{\textbf{Sci.\textbackslash F}}&\rot{\textbf{Hum.\textbackslash F}}&\rot{\textbf{Sci.\textbackslash M}}&\rot{\textbf{Hum.\textbackslash M}}&\rot{\textbf{Sum}}\\
\hline
$p<.05$& 1 & 3 & 3 & 2 & 3 & 0 & 1 & 5 & 2 & 20\\
\hline
$p<.01$& 1 & 1 & 4 & 1 & 1 & 0 & 1 & 1 & 0 & 10\\
\hline
$ p<.001$& 1 & 0 & 0 & 0 & 0 & 0 & 2 & 0 & 0 & 3\\
\hline
\textbf{Tot}& 3 & 4 & 7 & 3 & 4 & 0 & 4 & 6 & 2 & 33\\
\hline
\end{tabular}
\end{table}

\par Table \ref{tab:ects} shows that without considering any demographics, there is very little correlation between social media apps usage and CFU. There are three positive occurrences, i.e., $\bar S$ of SNS usage in general (-.26, $p<0.05$), $\bar I$ of IM apps while studying (0.29, $p<0.05$)and $\bar I$ of all the apps during lessons (0.28, $p<0.05$). Taking into account students' field of study, while for humanities the $\bar D$ parameter is related to a lower number of CFU, for scientific students $\bar S$ and $\bar I$ of SM apps tends to be more associated with increased CFU. Considering gender, SM negatively affect CFU from females especially during lessons: $\bar D$ of SM (-0.47, $p<0.05$) and SNS (-0.55,  $p<0.01$) plus $\bar I$ of browser apps (0.45, $p<0.05$); this effect is even stronger for females in humanities, with $\bar D$ of SM reaching -0.84 ($p<0.001$). For males, the pattern is less clear but the parameter that has a stronger negative correlation on CFU is $\bar S$ of SNS during general activities (-0.41, $p<0.01$).
\par Table \ref{tab:gpa} shows that, if we control for GPA without including demographics, we find the same trend of CFU results: $\bar I$ of all apps  while both studying (0.26, $p<0.05$) and attending lessons (0.40, $p<0.001$), and specifically $\bar I$ of Web apps during lessons, are positively associated with their GPA. Taking into account students' field of study, app usage significantly affects GPA while studying, with stronger effects for scientific students than humanities. Moreover, scientific students' GPA increases if they have higher $\bar I$ for all the apps (0.28, $p<0.05$) and for IM apps (0.30, $p<0.05$) and it decreases with higher level of  $\bar S$ for SNS apps (-0.41, $p<0.01$) while studying. Once again, also for GPA, the negative influence of social media app usage  for females occurs while attending lessons. Indeed, $\bar D$ of social media apps in general (-0.51, $p<0.05$) and of SNS in particular (-0.33, $p<0.01$) affects females performance especially while they are in the classroom. Overall, these findings suggest that:
\begin{itemize}
    \item There appear to be no major differences between males and female in terms of correlation of SM app usage and academic performance.
    \item Academic performance of scientific students is more affected by their SM usage than students from humanities. While this is an interesting finding, its causes are unclear and require further research.
    \item $\bar S$ and $\bar D$ are always correlated with lower performance both for GPA and for CFU while longer inactive periods ($\bar I$) are positively associated with them.
\end{itemize}


%% file: conclusions.tex
\section{Conclusions and Limitations}
\label{sec:conclusions}
In this paper, we proposed to overcome the current limitations of the state of the art in linking students' usage of social media on smartphones by coupling smartphones and time diaries, to then be able to match reports of time use with actual logs of SM apps. Based on the sample from the SmartUnitn project, we could corroborate the finding of sociological literature by using three parameters that pinpointed behavioural patterns that could either hurt academic performance, e.g., constantly messaging while studying or staying on SNS while in class, or improve it, e.g., limiting IM usage. 
\par Our results seems to confirm our hypothesis that controlling the use of social media during studying or lesson attending is more informative with respect to their association with students' academic performance than considering their general use without activity distinction, in addition to their faculties. 
\par Nonetheless, there two main limitations in our work. The first one is the relatively small window of time considered, i.e., two weeks, compared to other studies in computational social sciences, e.g., 10 weeks in SmartGPA \citep{wang2015smartgpa} and almost one year in the Copenhagen Networks Study \cite{karpinski2013exploration}. However, notice that one week of time diaries is considerably more than the usual amount of days recorded in sociology, which is usually limited to two days (one weekday and one weekend) \citep{romano2008time}, and thus allowed us a bigger time window to extract patterns from. The second one is the number of students in our sample. While it is considerably smaller than other studies in sociology, e.g., 263 students in \citep{rosen2013facebook} and 1839 students in \citep{junco2012too}, our sample is still is larger than other works in the area of computational social sciences, e.g., 48 students in SmartGPA \citep{wang2015smartgpa} and 35 students in \citep{leecomparing}. Although we are aware of the possible limitations of this study in terms of sample and window of time covered, they will be addressed in the next iteration of SmartUnitn to be carried out in March of 2018, by also detailing more the type of students' faculty, given how relevant it is in light of our results. 